\documentclass{elsart}

 \usepackage{graphicx}
  \include{epsfig}
\journal{Physica A}
\begin{document}
\begin{frontmatter}


\title{Evidence of Increment of Efficiency of the Mexican Stock Market Through the Analysis of its Variations}
\author{H.F. Coronel-Brizio$^\dagger$},
\ead{hcoronel@uv.mx}
\author{A.R. Hern\'andez-Montoya$^\dagger$},
\ead{alhernandez@uv.mx}
\author{R. Huerta-Quintanilla$^+$},
\ead{rhuerta@mda.cinvestav.mx}
\author{M. Rodr\'{\i}guez-Achach$^+$\corauthref{cor}}
\ead{achach@mda.cinvestav.mx}
\ead[url]{www.mda.cinvestav.mx}

\corauth[cor]{Corresponding author: Centro de Investigaci\'on y Estudios 
Avanzados (Cinvestav) del IPN, unidad M\'erida.}

\address{$^\dagger$ Facultad de F\'{\i}sica e Inteligencia Artificial.
Universidad Veracruzana, Sebasti\'an Camacho 5, Xalapa, Veracruz 91000. M\'{e}xico.}

\address{$^+$ Departamento de F\'{\i}sica Aplicada. Centro de Investigaci\'on y de Estudios Avanzados del IPN. Unidad M\'erida. Antigua carretera a Progreso km. 6, M\'erida, Yucat\'an, 97310, M\'exico.}

\begin{abstract}
  \noindent
It is well known that there exist statistical and structural differences 
between the stock markets of developed and emerging countries. In this work, 
we present an analysis of the variations and autocorrelations of the 
Mexican Stock Market index (IPC) for  different periods of its historical 
daily data, showing evidence that the Mexican Stock Market has been 
increasing its efficiency in recent times. We have analyzed the returns 
autocorrelation function (ACF) and used detrended fluctuation analysis 
(DFA) methods. We also analyze the volatility of the IPC and the Dow Jones
Industrial Average (DJIA) and compare their evolution. The data samples 
analyzed here, correspond to daily values 
of the IPC and DJIA for the period 10/30/1978 to 02/28/2006.
\end{abstract}

\end{frontmatter}



\section{Introduction}
\label{intro}
\vspace*{-.5cm}

\noindent
Several empirical studies of worldwide financial time series have  produced 
very interesting and important results under both  theoretical and 
practical points of view \cite{cons,pler,matos,grech,xin}. Examples are the 
determination of the power law character of the distribution of 
the variation of assets returns; temporal autocorrelations of returns decaying 
to zero in a few minutes following a random walk process and very long 
nonlinear correlations for their absolute returns (long term memory of the 
volatility of returns). 
These statistical properties are part of the so-called  ``stylized facts'' of 
financial markets \cite{sty_facts0,sty_facts}.
\noindent
Also, many studies have been carried to test the Efficient Market Hypothesis \cite{fama}. It establishes that at any given time, the prices of  traded assets fully and instantly reflect all available information reaching the market,
 and the market is said to be efficient. Nowadays it is clear that a completly efficient market is only an idealization and that different markets have different degrees of efficiency. However, more recent empirical studies seem to show that 
financial markets are evolving and increasing their efficiency 
over time \cite{toth,cajueiro}.

\noindent
It is well known that  emerging stock markets display different  structural 
and statistical properties\footnote{The later reflect the former} with regard 
to those belonging to developed countries. Among them, we can mention the 
following structural differences: higher sensitivity to capital flows, 
slower reaction to new information and a bigger effect of nonsynchronous 
trading on the prices. Among the statistical differences we have:  
emerging markets are more volatile, with higher average returns, slower 
autocorrelation function decrements and bigger Hurst 
exponents \cite{emerge1,emerge1_1,emerge2,faruk,matteo,beben}. All of the 
above can be described in economic terms saying that emerging markets  
are ``less efficient" than well established stock markets.
\noindent
In this work we use the Dow Jones Industrial Average (DJIA) index as a benchmark,  and study the autocorrelations and other statistical properties of the Mexican 
stock market, which can be considered as a relatively new\footnote{Even 
if some stock transactions started in M\'exico in 1895, the Mexican 
stock market only became well established, public and regulated until 1975.} 
and emerging stock market. Using the above techniques we are able to show 
that there is good evidence that the Mexican stock market has been increasing  
its  efficiency in recent years.

\noindent
This paper is organized as follows: In section \ref{section:acf}, an analysis 
of the ACF for different time periods of the IPC is presented and compared 
with those of the DJIA, showing that the Mexican index ACF time decay is 
becoming shorter and the amplitude of the ACF fluctuations smaller. In section \ref{section:DFA}, we applied the Detrended 
Fluctuation Analysis (DFA) statistical method  to the IPC and DJIA returns 
series. A comparative analysis gives a signal suggesting maturation and more 
efficient behavior of the IPC. In section \ref{vola},
a volatility analysis shows how the volatility of IPC has decreased steadily 
in time and particularly in the last 3 years has become comparable to that 
of the DJIA. Also in this section a trend analysis of both IPC and DJIA 
mean value and standard deviation is performed, showing that IPC variations 
evolve to tie those of the DJIA.
\noindent
Finally a summary of all results and conclusions derived from this work is 
presented in the last section of this paper.

\section{Autocorrelation Function of the IPC variations for different time
periods}
\label{section:acf}
\vspace*{-.5cm}
\noindent
In this study, we analyze the IPC and DJIA returns defined as: 
$S(t) = \ln P(t+\Delta t)-\ln P(t)$, 
where $P(t)$ is the value of the index at time $t$ and $\Delta t$ = 1 trading 
day. Since we lack high frequency data for the IPC, we have used daily 
values in our analysis. Despite this fact, we show that there 
is clear evidence of the evolution towards maturity and increment of 
efficiency of the IPC index, as is shown below.

\begin{figure}[htb!]
\begin{center}
\resizebox{0.65\textwidth}{!}{%
\includegraphics{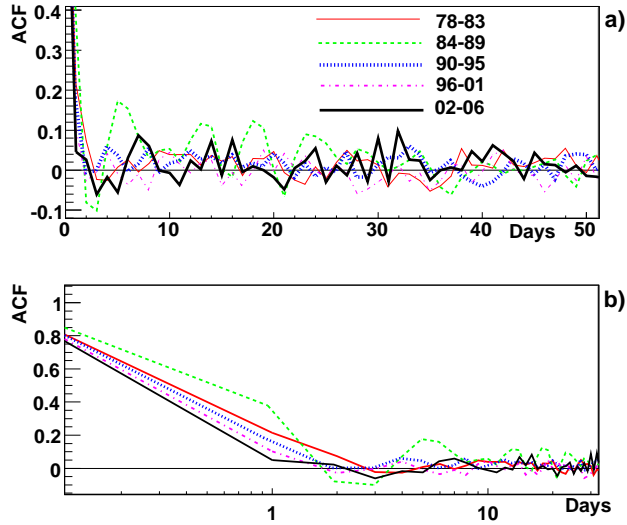}}
\caption{a) IPC daily returns ACF comparison for different periods. b) Same as before, but in a semi-log plot. It is possible to appreciate a tendency of the ACF to decay faster over time.}
\label{correla1}
\end{center}
\end{figure}

\begin{figure}[htb!]
\begin{center}
\resizebox{0.65\textwidth}{!}{%
\includegraphics{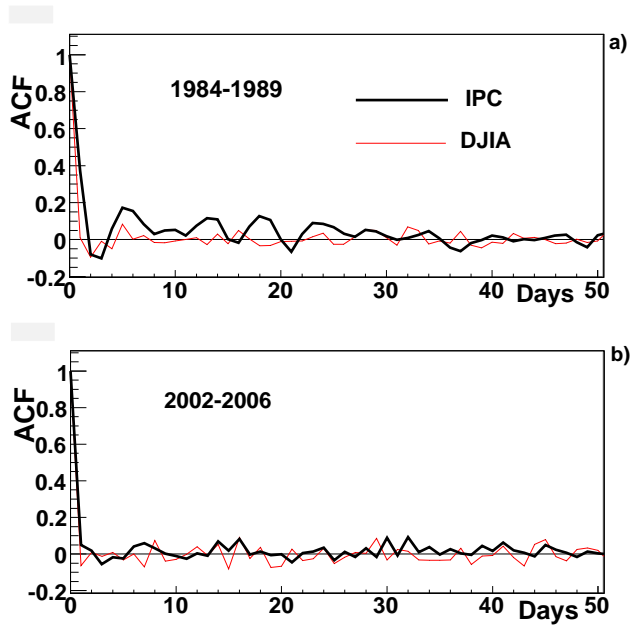}}
\caption{a) IPC and DJIA returns ACF. It is easy to appreciate that during the time period 1984-1989, IPC ACF decays slowly and has a bigger amplitude than 
that corresponding to the DJIA. b) The discrepancy between IPC and DJIA ACF 
in a) is no longer present in the period 2002-2006.}
\label{correla2}
\end{center}
\end{figure}

\noindent
The IPC data sample covers the period from 10/30/1978 to 02/28/2006 and was 
divided in 5 subperiods of about 6 years each: 1978-1983, 1984-1989, 
1990-1995, 1996-2001 and 2002-2006. We calculated the returns ACF for each one 
of these subperiods and show them in the figure \ref{correla1}a). In figure \ref{correla1}b) we can see the same data with a time-log scale. 
There is a clear tendency for the ACF to decrease faster over time.
\noindent
In figure \ref{correla2}a),  we compare the IPC returns ACF for 
the period 1984-1989 with the DJIA returns ACF for the same period. We can 
observe that the amplitude of the fluctuations of the IPC decays to a noise level more slowly when compared to the DJIA. Moreover, some kind of 
periodical amplitude fluctuations can be seen, with a period of about 8 days 
decaying to zero after about 25 days.
This behavior is not to be expected in an efficient market.
\noindent
Next we do the same comparison for data belonging to the most recent time 
period, i.e., for the period 2002-2006. This is shown in the lower panel of 
figure \ref{correla2}, where we can see that IPC and DJIA's ACFs simultaneously
reach the noise level.  Also, the fluctuations are very similar in both cases.
Information shown in figures \ref{correla1} and \ref{correla2} suggests 
that, from the point of view of the ACF, the Mexican stock market is becoming 
similar to the DJIA, that is, it is becoming more efficient.\\

\begin{figure}[htb!]
\begin{center}
\resizebox{0.65\textwidth}{!}{%
\includegraphics{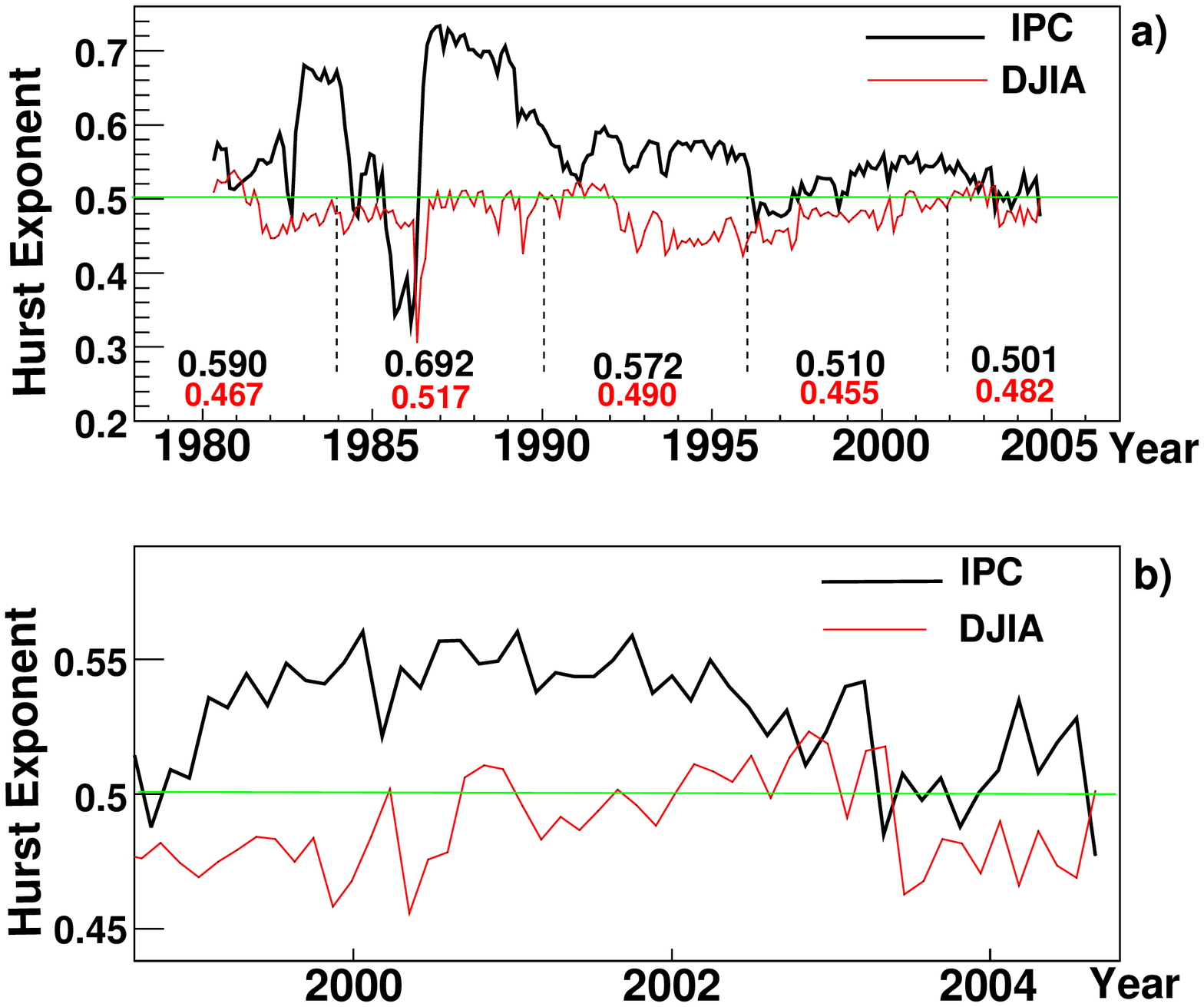}}
\caption{a) Values for the Hurst exponent over a 3 year time window for the
IPC and DJIA indexes. b) A close-up of the period 1999-present is shown
in the lower panel.}
\label{dfa_ipc_dji}
\end{center}
\end{figure}



\section{Detrended Fluctuation Analysis of the IPC}
\label{section:DFA}
\vspace*{-.5cm}
\noindent
Signal analysis using detrended fluctuation analysis was introduced
by Peng and collaborators \cite{peng_dfa}. It has proved to be a valuable
tool when searching for long time correlations in non-stationary time series.
In this method, a scaling or Hurst exponent $\alpha$ is computed for the time
series under analysis, whose values can be interpreted as: 
\begin{itemize}
\item i) if $\alpha=0.5$ there is no correlation at all and the time series represents
a random walk; 
\item ii) if $0.5<\alpha<1$, persistent long-range power-law 
correlations are present in the time series, and;  
\item iii) if $0<\alpha<0.5$, power-law anti-correlations are present.
\end{itemize}

\noindent
In the upper panel of figure \ref{dfa_ipc_dji}, we show the Hurst exponent
$\alpha$ computed for a time window of 3 years as a function of time, for both
the IPC and the DJIA. It can be seen how the values of $\alpha$
in the initial years of the IPC (1980-1990), are much more
erratic than those for DJIA. Great departures larger than a value of
 $\alpha=0.5$ is a characteristic of an emerging or non-mature
financial market \cite{beben}. In contrast, the DJIA, which can
be considered a mature market since decades ago, shows much less
pronounced variations in $\alpha$ for the same period with a value smaller than (but very) close to $0.5$.  

The six year averaged values for $\alpha$ shown in figure \ref{dfa_ipc_dji}a) show how this exponent
has been moving towards a value of 0.5 for the IPC, in particular
it has a value of $\alpha=0.51$ for the period 1996-2002, and
$\alpha=0.501$ for 2002-present. 
\noindent
Figure \ref{dfa_ipc_dji}b) shows a close-up of the
period 1999-present, where it is clear how the two indexes behave
in a similar fashion, both fluctuating around $\alpha=0.5$. 
Despite the relative small sizes of data samples analized 
\cite{finite_sample}, we believe the DFA analysis summarized in this figure 
stands as good evidence of the maturation of the IPC in recent times.

\section{Volatility Analysis}
\label{vola}
\vspace*{-.5cm}
Volatility is a term used for a statistical measure of value 
mean fluctuations over a certain interval of time of a given security or 
market index. A volatility measure still used in finance is the standard 
deviation of the security variations. However, because it is now well known 
that those changes do not follow a Gaussian distribution, other measures 
to describe these changes are preferred. In this section, volatility is 
calculated \cite{calc_vola} by taking the absolute returns and averaging them over a time 
window $T=n \Delta t$, i.e. 

\begin{center}
\begin{equation}
V(t):= \frac{1}{n}\Sigma_{t'=t}^{t+n-1}|S(t')|.
\end{equation}
\end{center}

\begin{figure}[htb!]
\begin{center}
\resizebox{0.65\textwidth}{!}{%
\includegraphics{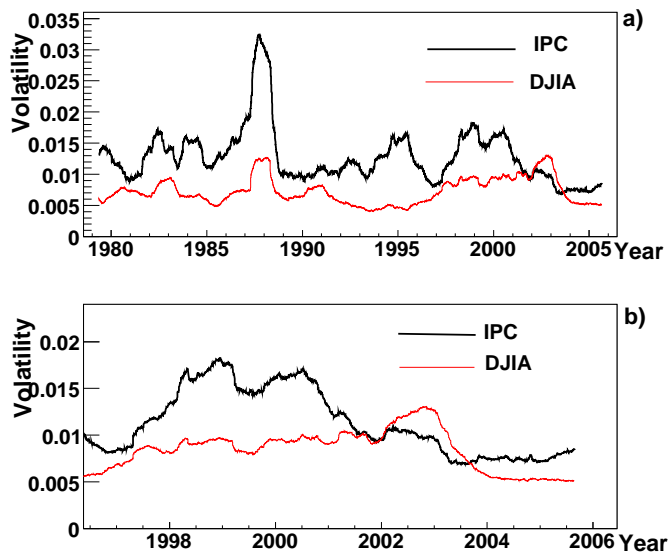}}
\caption{Comparison of the volatility in the IPC and DJIA indexes. Values of
volatility are computed for a 3 year time window. The lower panel is a 
zoom in of the region 1997-present.}
\label{correlavola}
\end{center}
\end{figure}

\noindent
In our case we used a $T=3$ years time window in order to compare IPC and DJIA 
volatilities and $\Delta t = $ 1 day. Both of them are displayed in figure 
\ref{correlavola}.

\noindent
The market's volatility can be used as an indicator for signs of maturation.  Persistent large values in the volatility are indicative 
of a market's immaturity. It is clear from  figure \ref{correlavola}a) that
the IPC's volatility was consistently higher than that of the DJIA, except
for more recent years (see Figure \ref{correlavola}b), where the two indexes show
fairly similar values of volatility. Again we believe this is a sign
of the evolution  of the Mexican Stock Market towards a more efficient behavior.

\subsection{Mean value and Standard Deviation analysis}
\begin{figure}[htb!]
\begin{center}
\resizebox{0.65\textwidth}{!}{%
\includegraphics{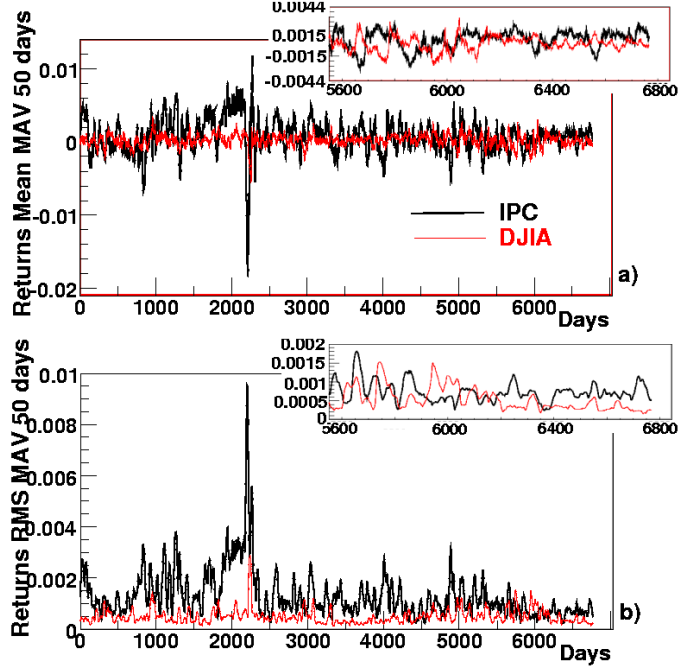}}
\caption{IPC and DJIA daily returns Mean (upper panel) and RMS (lower panel) evolution for the 10/30/1978 to 02/28/2006 period.}
\label{musigma}
\end{center}
\end{figure}

\noindent 
Finally, for completeness we have calculated and compared the 50 day 
Simple Moving Average (SMA) of the mean and standard deviation (RMS) of the 
IPC and DJIA returns for the total period 10/30/1978 to 02/28/2006. They are shown in figures \ref{musigma}a) and \ref{musigma}b).  

\noindent
It is interesting to observe that  the IPC returns fluctuations appear to  
become smaller in both mean value and RMS, i.e. seems that  IPC is becoming 
more well centered around zero and the amplitud of their fluctuations lessened. 
Trying to gain some insight and confirm this behavior, we have  calculated 
the returns mean value and RMS independently for each one of the periods 
already analyzed in previous sections. These values are shown in 
tables  \ref{tab:mu}  and \ref{tab:sigma}.

\noindent
From table \ref{tab:mu} is not easy to appreciate a clear trend of how the 
IPC and DJIA mean values of returns fluctuations are evolving over time; 
however, with respect to the standard deviation and observing the second 
column of table \ref{tab:sigma}, we can say that a clear decreasing trend 
in the IPC RMS exists, and therefore it is possible to say that the 
IPC is becoming less volatile, with a strong tendency to tie DJIA  
variations as can be appreciated from fourth column of the same table. 
This fact confirms again the evolution of the IPC to maturity.

\begin{table}
\begin{centering}
 \begin{tabular}{|c|c|c|c|c|}
 \hline
 Period& IPC $\mu$$(\times 10^{-4})$&DJIA $\mu$$(\times 10^{-4})$&
$\frac{IPC \mu}{DJIA \mu}$\\
 \hline
 \hline
 78-83 & $8.9\pm4.0 $& $3.4\pm1.9$&$\approx 2.62$\\
 84-89 & $34.7\pm4.8$& $5.2\pm2.3 $&$\approx 6.67$\\
 90-95 & $12.7\pm3.0 $ &$3.9\pm1.4$&$\approx 3.26$\\
 96-01 & $5.2\pm3.4$& $4.4\pm2.1 $&$\approx 1.18$\\
 02-06 & $10.3\pm2.4$& $0.8\pm2.3 $&$\approx 12.88$\\
\hline
 \hline
 \end{tabular}
 \caption{Returns mean values for the IPC and DJIA. We can not say if  a IPC tendency of its mean values to become close to zero exists }
 \label{tab:mu}
 \end{centering}
 \end{table}

 \begin{table}
 \begin{centering}
 \begin{tabular}{|c|c|c|c|c|}
 \hline
 Period&IPC $\sigma$$(\times 10^{-4})$&DJIA $\sigma$$(\times 10^{-4})$&
$\frac{IPC \sigma}{DJIA \sigma}$
\\
 \hline
 \hline
 78-83& $200.2\pm2.8 $&$94.6\pm 1.3$&$\approx 2.12$ \\
 84-89& $259.3\pm3.4$ &$26.0 \pm1.6$ &$\approx 9.97$\\
 90-95& $165.4\pm2.1 $ &$75.2 \pm 1.0$ &$\approx 2.20$\\
 96-01& $183.6\pm$2.4 &$116.1 \pm1.5 $ &$\approx 1.58$\\
 02-06& $ 109.2\pm1.7$ &$105.3 \pm1.6$ &$\approx 1.04$\\
\hline
 \hline
 \end{tabular}
 \caption{Returns standard deviation for the IPC and DJIA for different time intervals. It is clear that IPC variations decrease and they tend to tie DJIA variations. Again, we are only interested in  these parameters trend behavior. 
}
 \label{tab:sigma}
 \end{centering}
 \end{table}

\section{Resume}
\label{resume}
\vspace*{-.5cm}
\noindent
We have used four distinct methods, namely: autocorrelation analysis, detrended fluctuation analysis, volatility analysis and  
standard deviation trending analysis, to study the maturation process of the 
Mexican stock market through its leading index, the IPC, and compared the 
results with a well developed and mature index, the Dow Jones Industrial 
Average (DJIA).
In all four cases, we have found compelling evidence of evolution
of the Mexican market towards a more efficient market. In particular, from 
around year 2000 to date, the four methods used in this study show 
that the IPC has remarkably similar behavior to that of the DJIA.

{\bf Acknowledgments}

\noindent
We thank the valuable suggestions from S. Jim\'enez, P. Giubellino,
A. Guerra and R. Vilalta. We also thank  P. Zorrilla-Velasco, A. 
Reynoso del Valle and S. Herrera-Montiel from the BMV for their valuable time and 
cooperation providing us with the IPC index daily data set.\\ 
\noindent
This work has been supported by Conacyt-Mexico under Grants 44598
and 45782.
\noindent
 Analysis has been realized using ROOT \cite{root}.

\end{document}